\documentclass[conference]{IEEEtran}
\IEEEoverridecommandlockouts
\usepackage{cite}
\usepackage{amsmath,amssymb,amsfonts}
\usepackage{algorithmic}
\usepackage{graphicx}
\usepackage{textcomp}
\usepackage{subfig}
\usepackage{xcolor}
\usepackage[table]{xcolor} 
\usepackage{academicons} 
\usepackage{hyperref} 
\definecolor{lightgreen}{RGB}{204, 255, 204} 
\newcommand{\orcidicon}{\includegraphics[height=10pt]{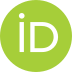}}
\newcommand{\orcidlink}[1]{\href{https://orcid.org/#1}{\orcidicon}}

\def\BibTeX{{\rm B\kern-.05em{\sc i\kern-.025em b}\kern-.08em
    T\kern-.1667em\lower.7ex\hbox{E}\kern-.125emX}}
\begin{document}

\title{Power Side-Channel Analysis of the CVA6 RISC-V Core at the RTL Level Using VeriSide
}

\author{\IEEEauthorblockN{Behnam Farnaghinejad \orcidlink{0009-0000-8456-3166}, Antonio Porsia \orcidlink{0009-0009-4671-5064}, Annachiara Ruospo \orcidlink{0000-0003-2040-9762}, \\ Alessandro Savino  \orcidlink{0000-0003-0529-7950}, Stefano Di Carlo \orcidlink{0000-0002-7512-5356} and Ernesto Sanchez \orcidlink{0000-0002-7042-295X}}
\IEEEauthorblockA{\textit{Department of Control and Computer Engineering, Politecnico di Torino, Turin, Italy} 
\thanks{This work was supported by project SERICS (PE00000014) under the MUR National Recovery and Resilience Plan funded by the European Union - NextGenerationEU.}
}
}
\maketitle

\begin{abstract}
Security in modern RISC-V processors demands
more than functional correctness: It requires resilience to side-channel attacks. This paper evaluates the vulnerability of the side channel of the CVA6 RISC-V core by analyzing software-based
AES encryption uses an RTL-level power profiling framework called VeriSide. This work represents that this design's Correlation Power Analysis (CPA) reveals significant leakage, enabling key recovery. 
These findings underscore the importance of early-stage RTL assessments in shaping future secure RISC-V designs.

\end{abstract}



\section{Introduction}

RISC-V processors are increasingly employed in security-critical applications, making resistance to side-channel attacks (SCAs) essential. Power-based SCAs (PSCAs) exploit power consumption variations during cryptographic operations to infer secret keys.

While previous studies have examined side-channel vulnerabilities in RISC-V processors, none have evaluated CVA6 \cite{b1}, a 64-bit Linux-capable processor highly suited to security-critical workloads.

This work focuses on RTL-level analysis of CVA6. Traditional RTL-level side-channel analyses often rely on waveform-based methods such as Value Change Dump (VCD) and Switching Activity Interchange Format (SAIF) files. However, their significant overhead makes these methods impractical for analyzing complex cores. The optimized RTL simulation framework VeriSide \cite{b0} was developed to overcome this limitation. VeriSide efficiently extracts power traces, represented as the Hamming distances or Hamming weights of signals, immediately after simulation without the additional overhead of waveform files.

This paper presents the first side-channel evaluation of CVA6, specifically targeting software-based AES encryption leakage. The results demonstrate that the CVA6 processor is susceptible to PSCA attacks.

Our scalable, technology-independent methodology enables effective early-stage RTL-level side-channel evaluations, providing critical insights for securing future RISC-V architectures.

\section{Background}

RISC-V is an open-source Instruction Set Architecture (ISA) celebrated for its flexibility and extensibility. CVA6 (formerly Ariane) is a 64-bit, Linux-capable application-class RISC-V core optimized for high-performance computing. It features configurable size, separate instruction, and data TLBs, a hardware Page Table Walker (PTW), and an advanced branch prediction unit with a Branch Target Buffer (BTB) and Branch History Table (BHT). Designed to minimize critical path length, CVA6 supports multiple privilege levels (M, S, and U), enabling the execution of Unix-like operating systems\cite{b1}.

Side-channel attacks exploit the physical characteristics of a processor, such as power consumption, to extract sensitive information. Techniques like masking and hiding can help mitigate these attacks, but identifying vulnerabilities in the early stage of design is crucial. Pre-silicon analysis at the Register-Transfer Level (RTL) allows designers to pinpoint potential leakage sources before fabrication, thus reducing costly hardware revisions later.

In CMOS-based digital circuits, dynamic power consumption constitutes a significant portion of the total power consumed. This dynamic power is primarily caused by transistors' switching activity (SA) as the digital signals transition between logical states (0 and 1). Mathematically, dynamic power consumption can be expressed by the well-known equation:

\begin{equation}
    P_{dynamic} = \alpha \cdot C_{L} \cdot V_{DD}^{2} \cdot f
\end{equation}

where:

\begin{itemize}
    \item $\alpha$ is the SA factor, representing the average number of signal transitions per clock cycle.
    \item $C_{L}$ is the load capacitance being driven by the gate.
    \item $V_{DD}$ is the supply voltage.
    \item $f$ is the frequency of operation.
\end{itemize}

Among these parameters, the SA factor ($\alpha$) directly links signal transitions to dynamic power consumption. A higher SA corresponds directly to increased energy dissipation due to repeated charging and discharging of load capacitances. Therefore, measuring or estimating switching activity provides a practical and effective approximation for assessing the power consumption of digital systems. 

In side-channel analysis contexts, such as correlation power analysis (CPA), switching activity thus serves as a reliable proxy for estimating power consumption traces, enabling attackers to correlate power measurements to internal circuit activities and secret data.

VeriSide \cite{b0} is a modified Verilator-based framework that computes Hamming Distance (HD) and Hamming Weight (HW) values directly during simulation, eliminating the need for large waveform files (e.g., VCD or SAIF). As shown in Table~\ref{tab:performance_rv64imfdc}, VeriSide notably reduces disk usage and post-simulation processing compared to standard Verilator \cite{b2} workflows, which rely on VCD file parsing (either storing all signal transitions before calculating HD or storing HD on-the-fly). Under the RV64IMFDC configuration with 98,523 signals, for example, the C program requires 283,858 clock cycles to complete 1,000 monitored function calls (approximately 150 cycles per call). This work uses VeriSide to extract CVA6 power data during simulation, providing a foundation for subsequent side-channel evaluations.

\begin{table}[htbp]

\caption{Performance Comparison of VeriSide}
\centering
\renewcommand{\arraystretch}{1.2} 
\resizebox{0.45\textwidth}{!}{%
\begin{tabular}{|p{1.8cm}|p{1.3cm}|p{2.2cm}|p{2cm}|}
\hline
\rowcolor{lightgray}\textbf{Metric} & \textbf{VeriSide} & \textbf{Verilator (Norm.)} & \textbf{Verilator (Opt.)} \\ \hline
\rowcolor{yellow}\multicolumn{4}{|c|}{Simulation} \\ \hline
CPU Time (ms) & 109,464 & 121,395 & 121,395 \\ \hline
Wall Time (ms) & 109,487 & 141,176 & 141,176 \\ \hline
Disk Usage & 3.8 MB & 5.55 GB & 5.55 GB \\ \hline
\rowcolor{lightgreen}\multicolumn{4}{|c|}{After Simulation (Parsing VCD file and extract HD/HW)} \\ \hline
Time (s) & --- & 753.00 & 588.00 \\ \hline
RAM Usage & --- & 46 GB & 14 GB \\ \hline
\end{tabular}}
\label{tab:performance_rv64imfdc}
\vspace*{-10pt}
\end{table}

\section{Methodology}

\subsection{Experimental Setup}
\begin{figure}[htbp]

    \centering
    \includegraphics[width=0.95\linewidth]{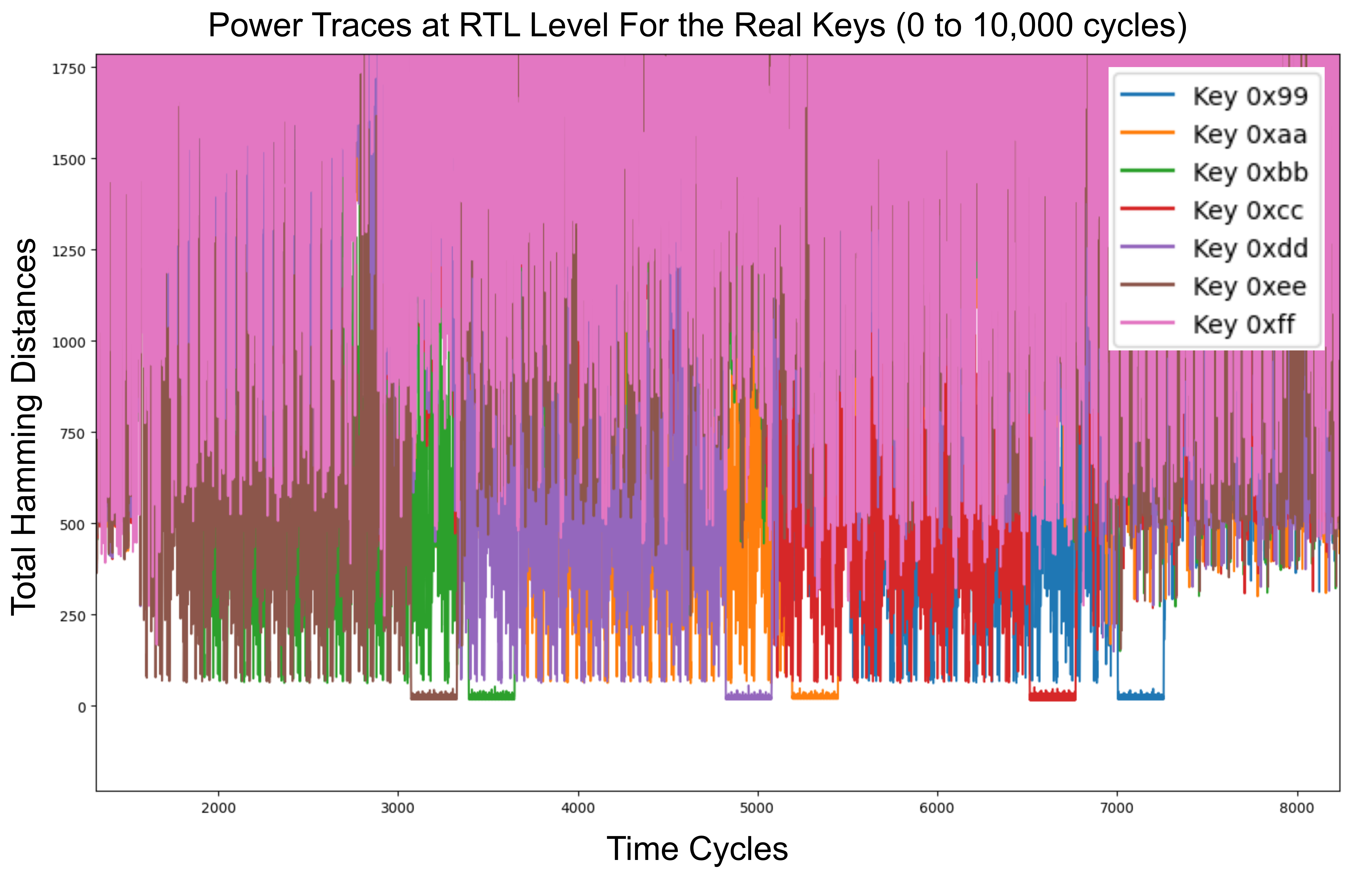}
    \caption{Power Traces (SA) of the Actual Keys}
    \label{fig:real_keys_zoom}

\end{figure}

A C program was developed to perform AES encryption on the CVA6 processor, employing a fixed encryption key \texttt{00ff00ff11ee22dd33cc44bb55aa6699}. The input plaintexts were structured by assigning sequential values (\texttt{0} to \texttt{3000}) to each 16-bit portion within the 128-bit plaintext. This selection strategy ensures that the first byte of each 16-bit portion contains a wide distribution, improving key recovery success.
Power traces were collected from RTL simulations using VeriSide, and 3000 encryption operations were recorded. The goal was to evaluate the power consumption variations and identify potential vulnerabilities in the AES execution on CVA6. Figure \ref{fig:real_keys_zoom} shows precise total hamming distances corresponding to key values over time.

\subsection{Correlation Power Analysis (CPA)}
We performed CPA using the Pearson correlation coefficient to evaluate the side-channel leakage of CVA6 running software AES. The correlation formula is given by:

\begin{equation}
\rho(H, T) = \frac{\sum (H_i - \bar{H})(T_i - \bar{T})}{\sqrt{\sum (H_i - \bar{H})^2} \sqrt{\sum (T_i - \bar{T})^2}}
\end{equation}

Where:
- \( H \) represents the hypothetical power model values based on key guesses.
- \( T \) represents the measured power traces extracted using VeriSide.
- \( \bar{H} \) and \( \bar{T} \) are the respective mean values.



\begin{figure}[htbp]

    \centering
    \includegraphics[width=0.9\linewidth]{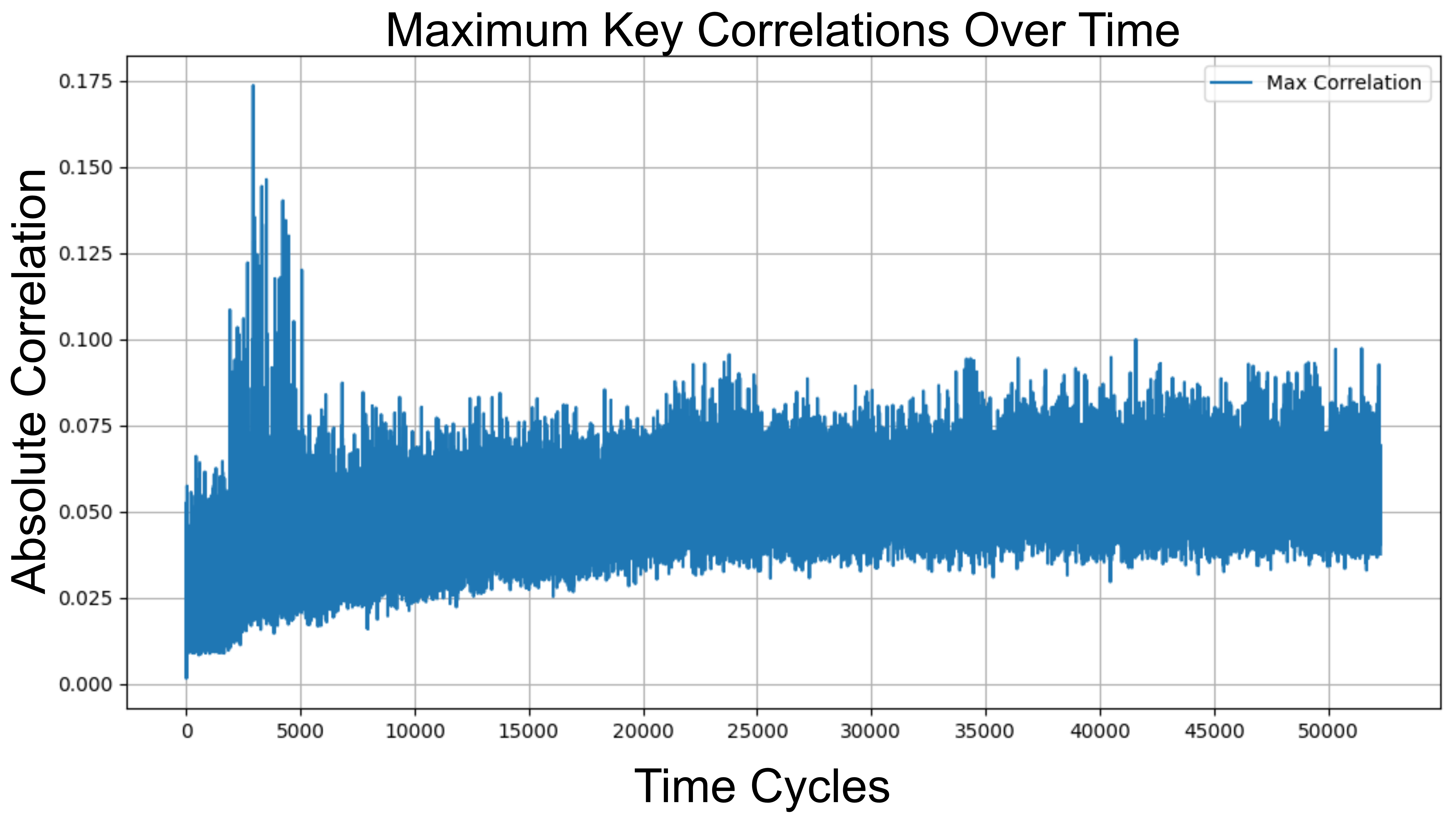}
    \caption{Maximum Correlation Over Time}
    \label{fig:max_corr}

\end{figure}

CPA was applied across all key guesses within a time window of 0 to 10,000 clock cycles, explicitly targeting the initial SBox operation of the AES encryption. Due to the structured plaintext selection, only the first byte of the key could be reliably extracted, with successfully recovered key bytes: \texttt{99, aa, bb, cc, dd, ee, ff}. Figure \ref{fig:max_corr} highlights peak correlation values, demonstrating successful identification of key leakage.

\subsection{Impact of Plaintext Distribution on CPA Results}

Plaintext distribution significantly affects CPA effectiveness. Structured plaintexts—formed by embedding repeated 16-bit patterns within each 128-bit block—yielded higher correlations primarily for the first byte of each segment, thereby limiting key extraction. Thus, careful plaintext selection is critical to reliably extracting secret keys.

\section{Discussion and Future Work}
This study exposed the side-channel vulnerability of CVA6 running software-based AES, highlighting the necessity for robust hardware countermeasures. Future research will extend the analysis to the RISC-V scalar cryptographic exstension\cite{b3}, assessing its security, enhancing its resilience to side-channel attacks, and exploring effective countermeasures.

\end{document}